\documentclass[12pt]{iopart}

\usepackage{graphicx}
\begin{document} 

\title[Elliptic and triangular flow of identified particles at ALICE.]
{Elliptic and triangular flow of identified particles measured with ALICE detector at the LHC.}

\author{Miko{\l}aj Krzewicki for the ALICE Collaboration.}
\address{NIKHEF,
Science Park 105,
1098 XG Amsterdam,
The Netherlands.}
\ead{mikolaj.krzewicki@cern.ch}

\begin{abstract}
We report on the first measurements of elliptic and triangular flow for charged
pions, kaons and anti-protons in lead-lead collisions at $\sqrt{s_{NN}} = 2.76\,\mathrm{TeV}$
measured with the ALICE detector at the LHC.
We compare the observed mass splitting of differential elliptic flow at LHC
energies to RHIC measurements at lower energies and theory predictions.
We test the quark coalescence picture with the quark number scaling of 
elliptic and triangular flow.
\end{abstract}


\vspace{-1 cm}

\section{Introduction}
Anisotropic flow \cite{ref:OllitraultFlow}, described by the coefficients in the Fourier expansion of the azimuthal
particle distribution \cite{ref:VoloshinZhang}, is an important probe of collectivity in the system
created in heavy ion collisions. The first measurement of elliptic 
flow ($v_2$) at the LHC shows an increase (with respect to RHIC energies) of the 
integrated value by about 30\% and no significant increase of the
$p_t$ differential flow \cite{ref:AalmodtFlow1st}. Hydrodynamical models \cite{ref:ShenHeinz} predict 
a rise of the radial expansion velocity from RHIC to the LHC energies which might 
explain this behaviour.
Elliptic flow of identified hadrons is sensitive to the hydrodynamical 
radial expansion of the medium. Since flow develops into a common velocity field
a mass-dependent shift towards higher momenta occurs giving
rise to mass splitting and ordering of flow for different 
particle species.
Triangular flow ($v_3$) of identified particles in hydrodynamics is expected to exhibit similar
mass scaling and additionally it is expected to be
a sensitive probe for the viscosity to entropy ratio $\eta/s$ of the created medium
\cite{ref:AlverHydroV3}. The measurement of identified particle $v_3$ can therefore
provide a constraint on $\eta/s$ after more detailed theoretical 
calculations become available for LHC energies. 

We present the measurement of identified particle $v_2$, compare it to the RHIC measurement
and test the quark number scaling in terms of transverse kinetic energy ($KE_t$ scaling). 
For identified particle $v_3$ we present the first preliminary results.

\section{Data analysis}
For this analysis the standard ALICE minimum bias event selection was used together
with an additional requirement on the primary vertex position $|z|<7\,\mathrm{cm}$ yielding 
a sample of around 4 million events.
The collision centrality was determined using the forward VZERO scintillator arrays \cite{ref:Toia}. 
Particle tracking was done
using the time projection chamber (TPC) and the silicon inner tracking system with full azimuth coverage for
$|\eta|<0.8$. The particle identification was done by combining the time-of-flight
(TOF) measurement with the energy loss measurement in the TPC. The purity
was estimated to be better than $95\%$ at $p<3\,\mathrm{GeV}/c$ for pions and kaons and at
$p<5\,\mathrm{GeV}/c$ for protons. In order to reduce the contamination from non-primary
particles the reconstructed particles were required to have a distance of closest approach to
the primary vertex of less than 1\,mm.
Main sources of systematic uncertainty on the flow values
considered in this analysis are non-flow, feed-down and centrality determination. In the following figures,
uncertainty bands indicate systematic and statistical uncertainties added in quadrature.
Elliptic flow is measured using the two-particle scalar product method \cite{ref:Adler2002} with a 
large $\eta$ gap ($|\Delta\eta|>1$) to reduce the contribution from short range
non-flow correlations.
\vspace{-0.3 cm}
\section{Results}
\begin{figure}[h!]
\includegraphics[width=0.5\textwidth]{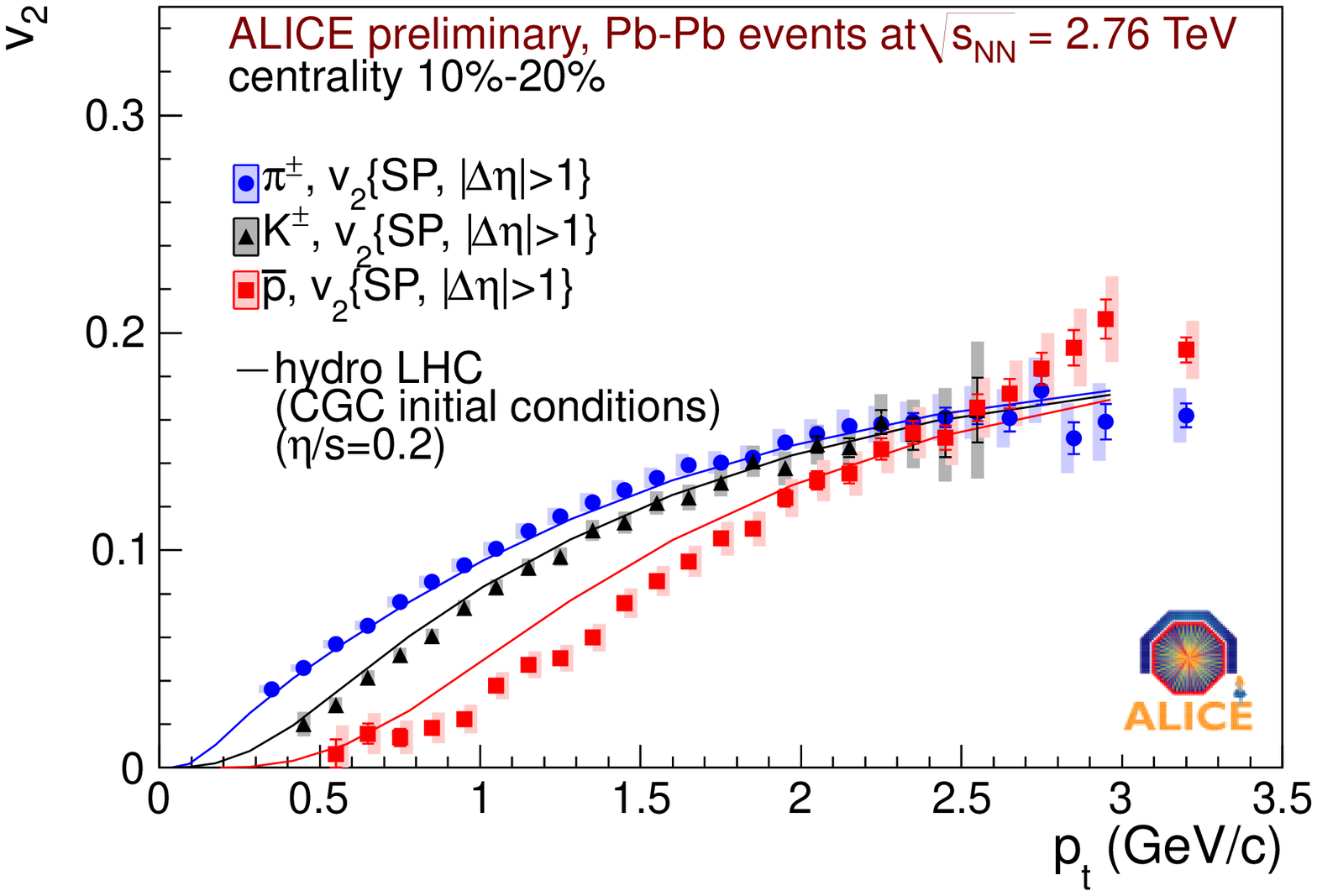}
\includegraphics[width=0.5\textwidth]{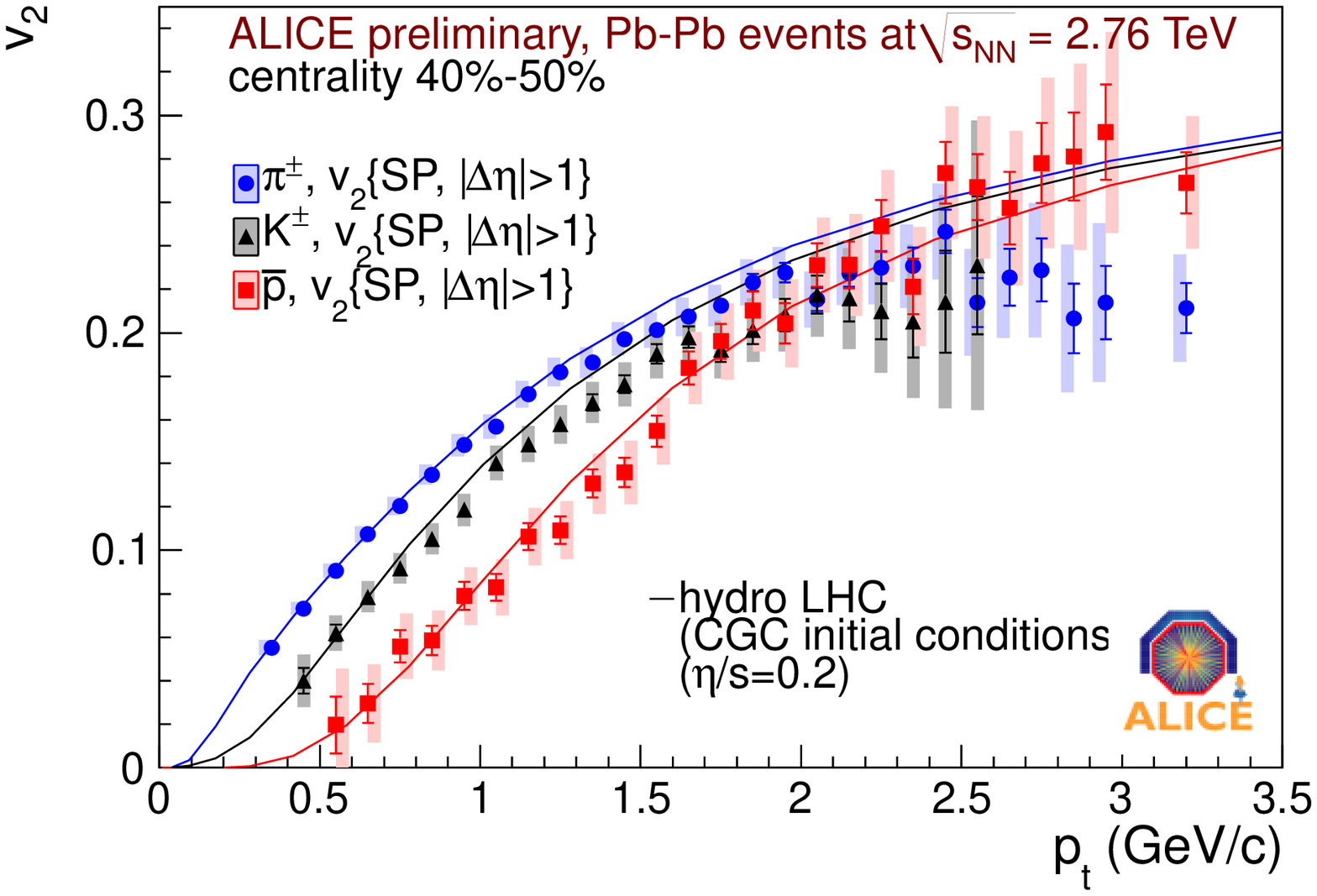}
\caption{Identified particle elliptic flow in two centrality bins compared to a hydrodynamic prediction
\cite{ref:ShenHeinz}.
Theory describes the data well at low to intermediate $p_{t}$ except for the anti-proton $v_2$ in the more 
central collisions (10\%-20\%).}
\label{fig:figure1} 
\end{figure}
\begin{figure}[h!]
\includegraphics[width=0.5\textwidth]{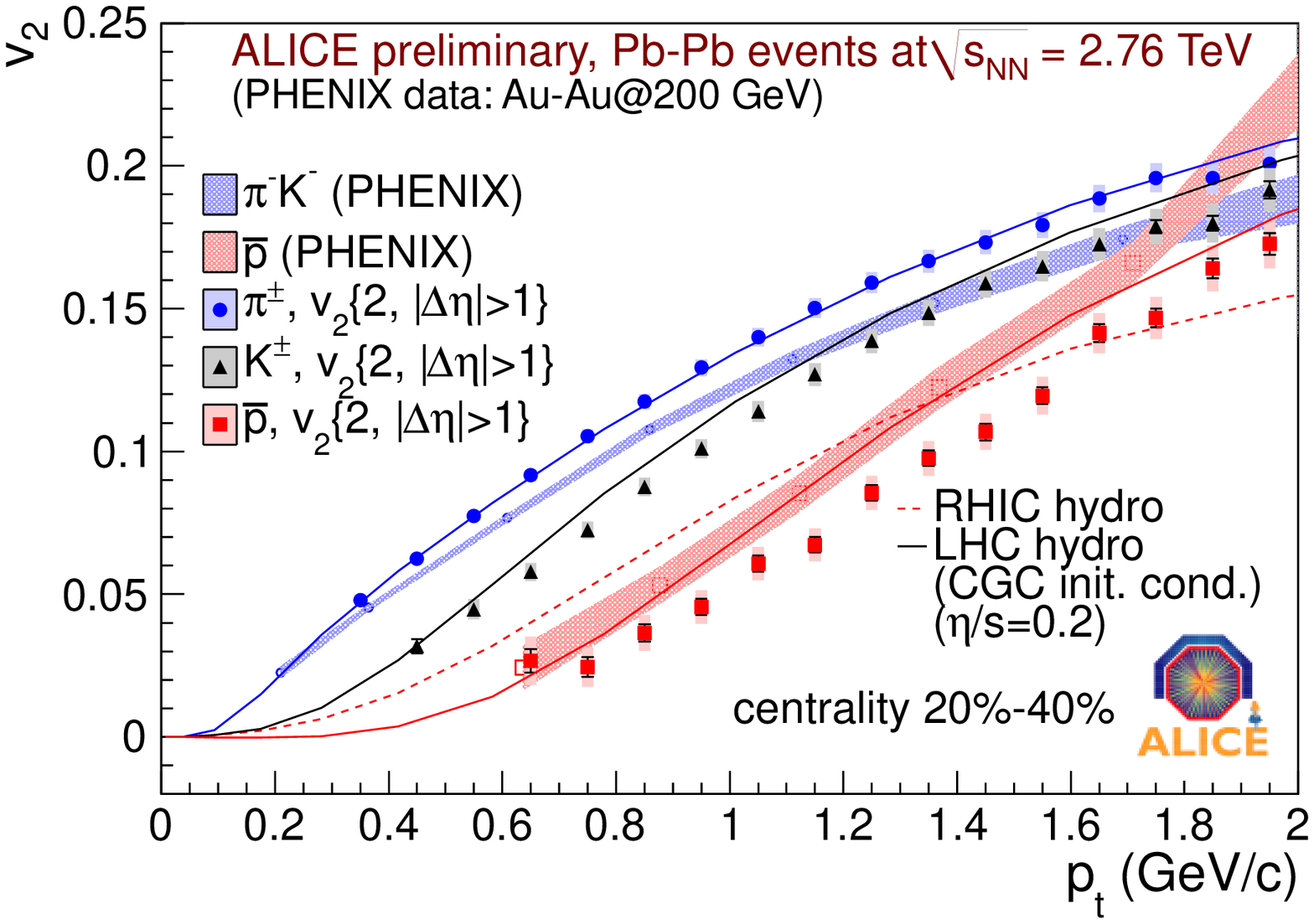}
\includegraphics[width=0.5\textwidth]{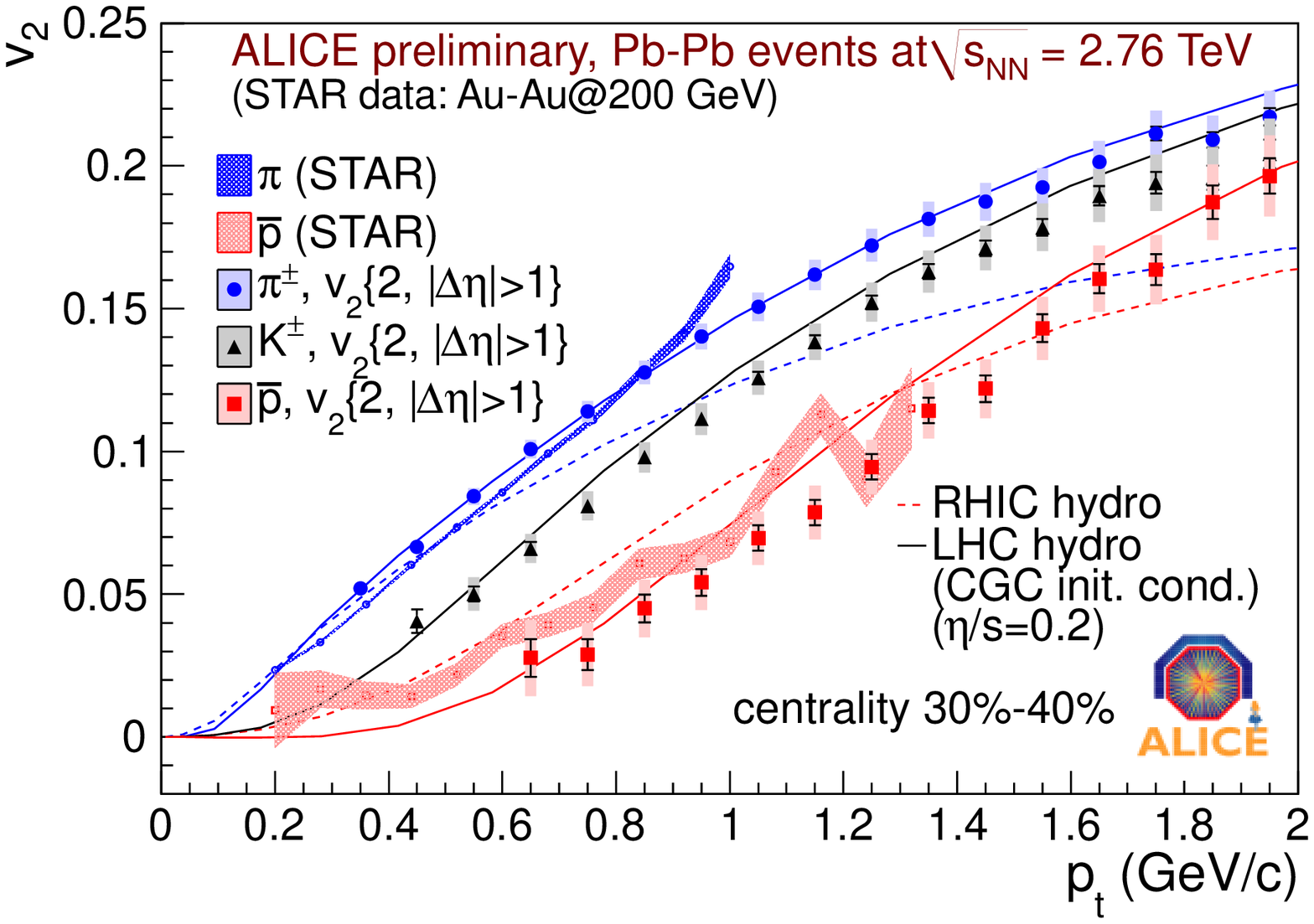}
\caption{Elliptic flow for Pb-Pb at $\sqrt{s_{NN}}=2.76\,\mathrm{TeV}$ 
compared to RHIC results shows a larger mass splitting than at RHIC energies.}
\label{fig:figure2} 
\end{figure}

Figure \ref{fig:figure1} shows elliptic flow measured for more central
($10\%-20\%$) and more peripheral ($40\%-50\%$) collisions overlaying a theoretical prediction from
ideal hydrodynamics \cite{ref:ShenHeinz}.
As can be seen, the theory is in agreement with the measurement for pions and kaons,
but in the case of the more central events this model does not describe anti-protons.
The measured $v_2$ is compared to the published results from the PHENIX \cite{ref:PHENIXdata}
and STAR \cite{ref:STARdata}
collaborations in figure \ref{fig:figure2}. Since PHENIX reports a combined result 
for pions and kaons we only compare the anti-proton results which show lower
values in ALICE data consistent with larger radial flow at $2.76\,\mathrm{TeV}$. 
In the comparison with STAR data we observe a larger mass splitting in both pion and anti-proton comparisons.

\begin{figure}[h!]
\includegraphics[width=0.5\textwidth]{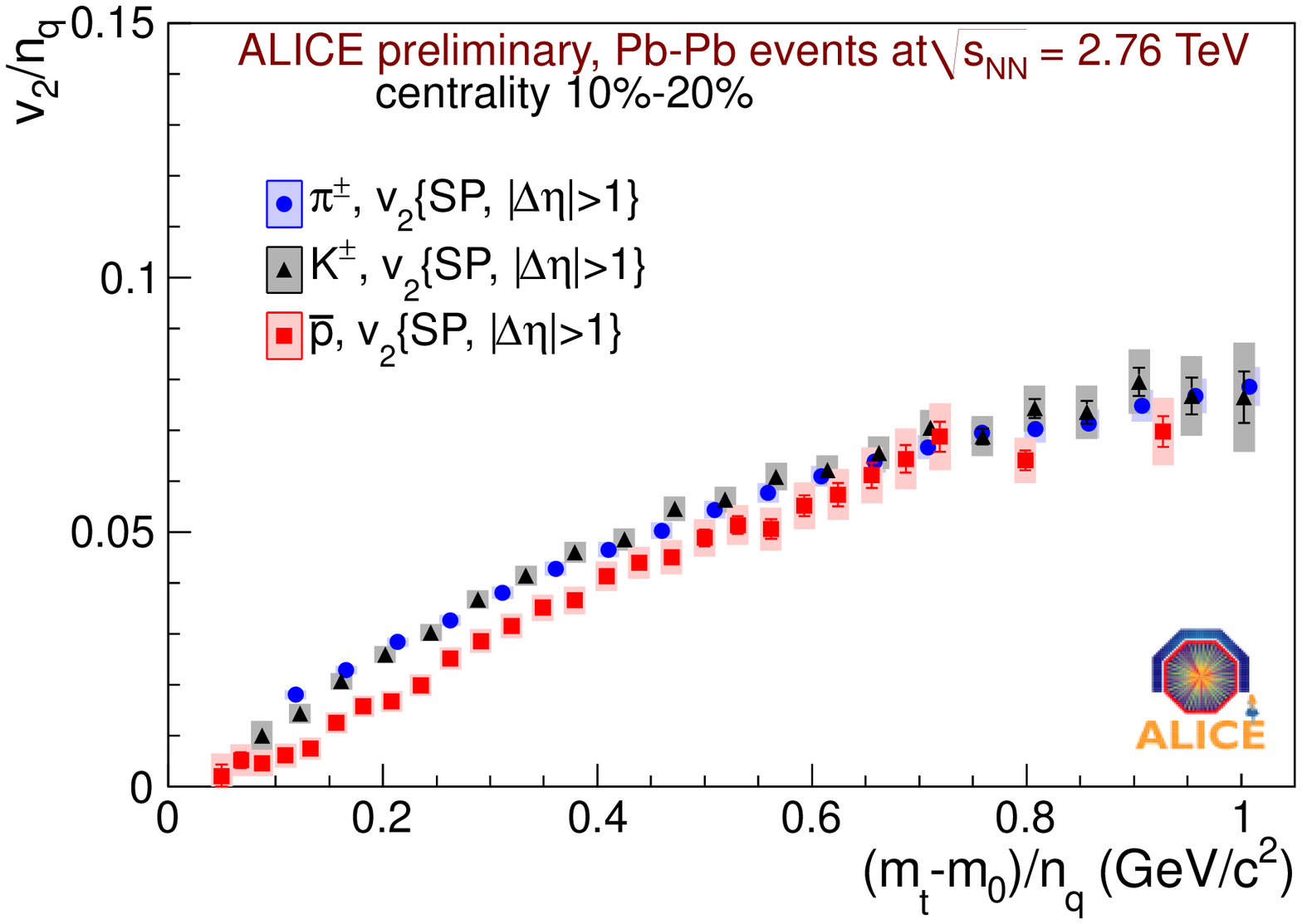}
\includegraphics[width=0.5\textwidth]{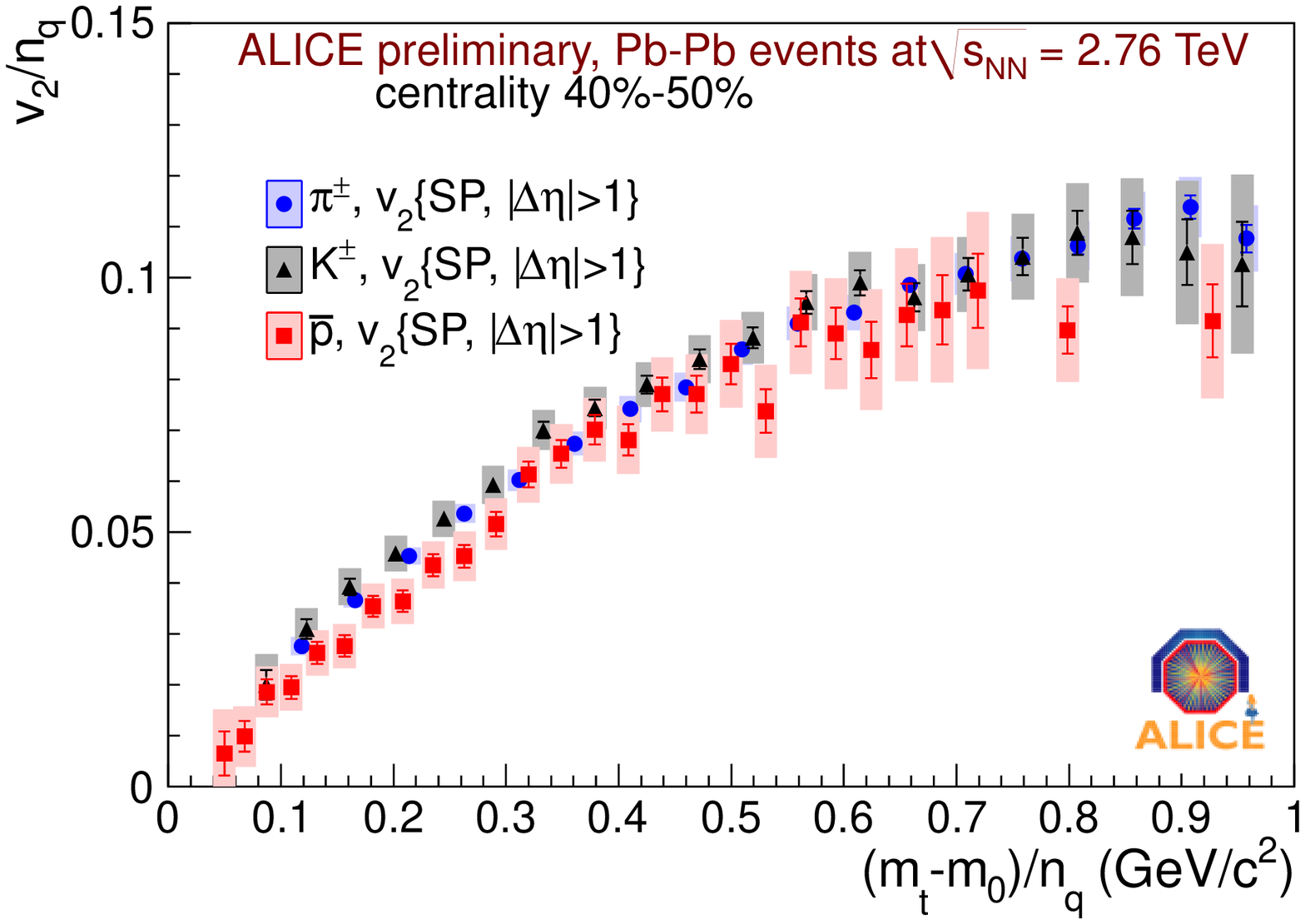}
\caption{Elliptic flow per constituent quark vs.\ transverse kinetic energy per quark
(the $KE_t$ scaling) for more central ($10\%-20\%$) 
and more peripheral ($40\%-50\%$) Pb-Pb collisions at $\sqrt{s_{NN}}=2.76\,\mathrm{TeV}$. }
\label{fig:figure3} 
\end{figure}

We report on the results at LHC energies of $KE_t$ scaling, introduced
and described in \cite{ref:AdareScaling} and \cite{ref:Taranenko}. 
From figure \ref{fig:figure3} it can be seen that, within errors, the flow 
of pions and kaons follows the scaling while the flow of anti-protons 
deviates for the more central and the more peripheral events.

\begin{figure}[h!]
\includegraphics[width=0.5\textwidth]{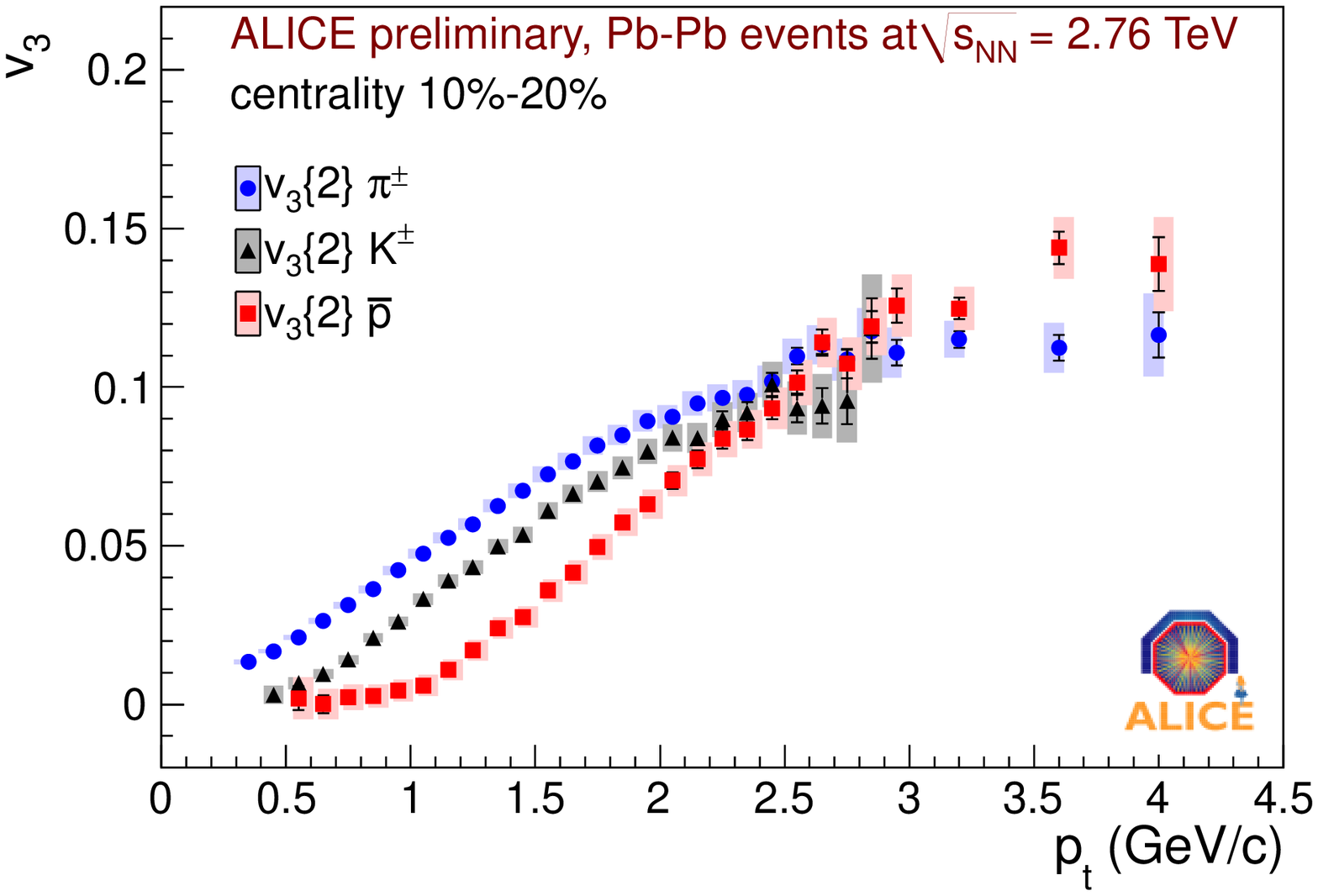}
\includegraphics[width=0.5\textwidth]{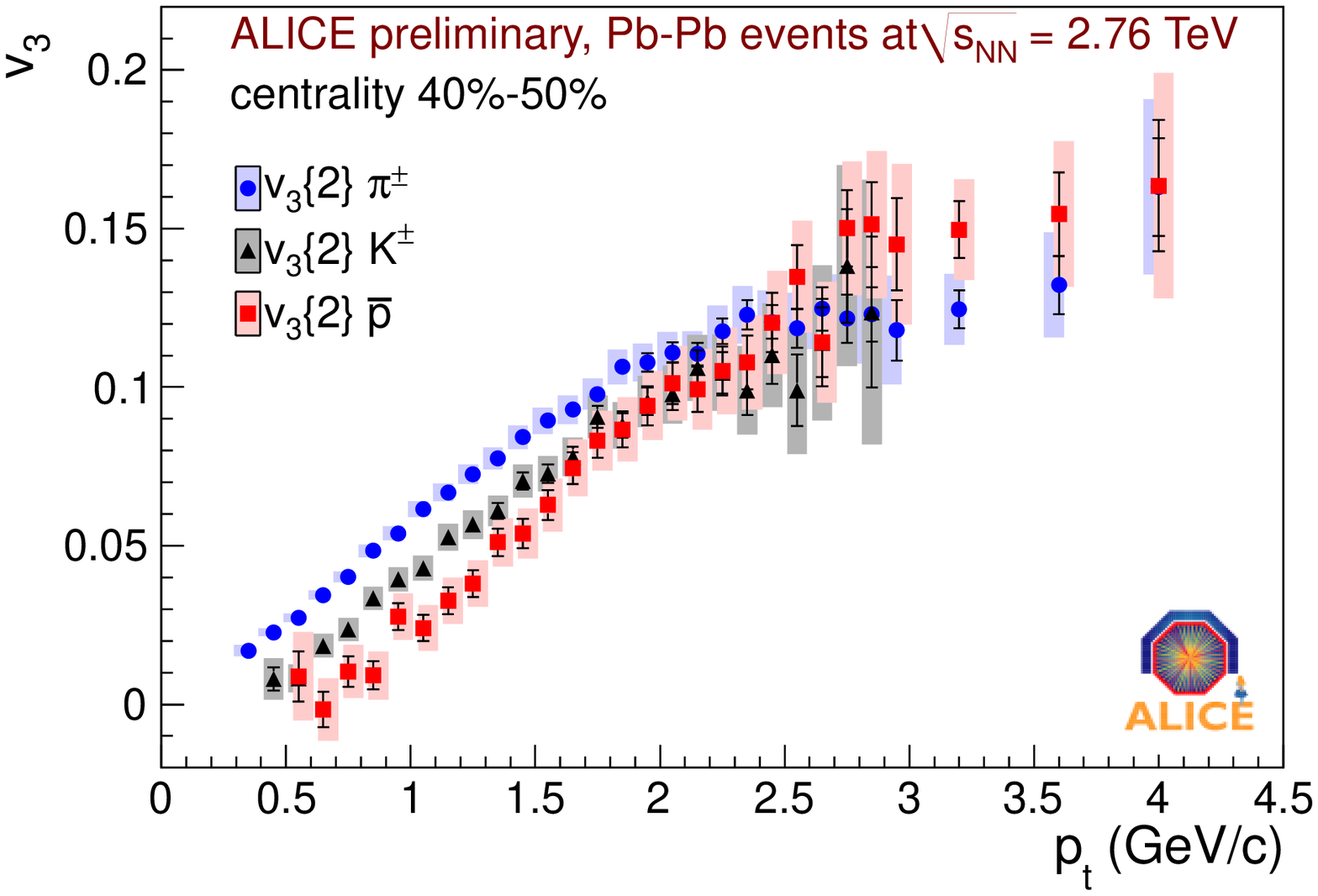}
\caption{Triangular flow for more central ($10\%-20\%$) 
and more peripheral ($40\%-50\%$) Pb-Pb collisions at $\sqrt{s_{NN}}=2.76\,\mathrm{TeV}$.}
\label{fig:figure4} 
\end{figure}

\begin{figure}[h!]
\includegraphics[width=0.5\textwidth]{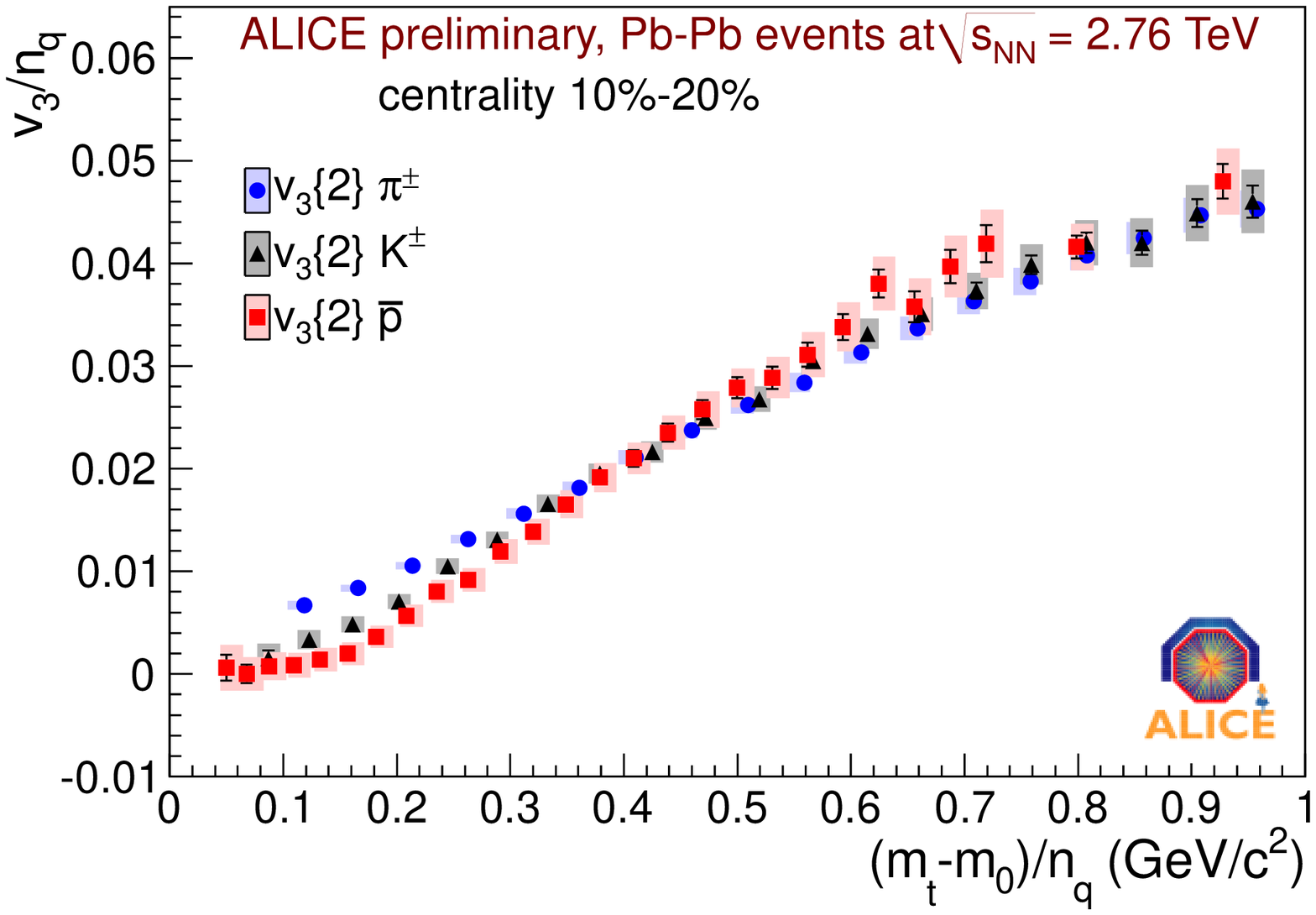}
\includegraphics[width=0.5\textwidth]{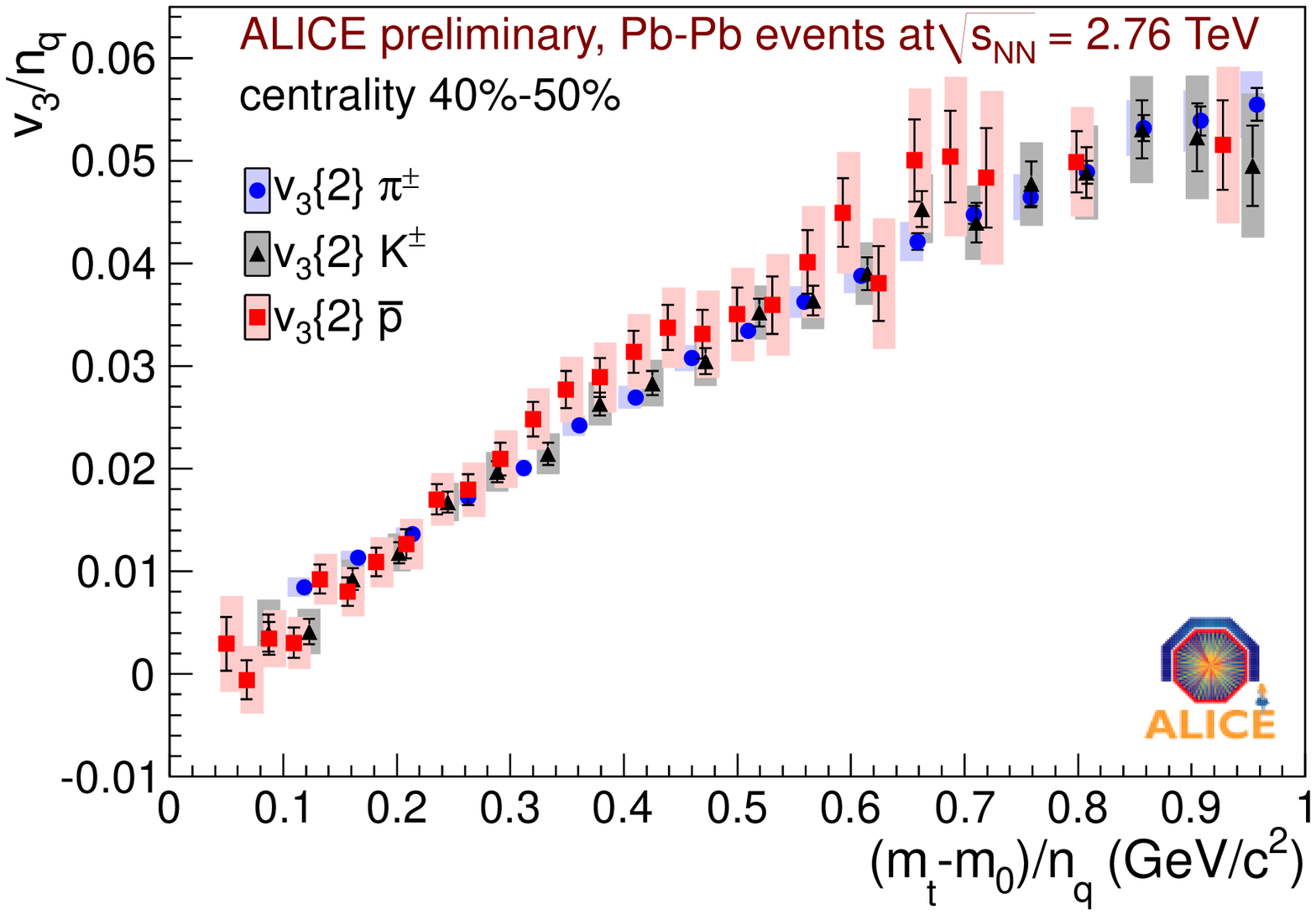}
\caption{$KE_t$ scaling of triangular flow for more central ($10\%-20\%$) 
and more peripheral ($40\%-50\%$) Pb-Pb collisions at $\sqrt{s_{NN}}=2.76\,\mathrm{TeV}$.}
\label{fig:figure5} 
\end{figure}

Triangular flow, depicted in figure \ref{fig:figure4}, qualitatively exhibits the
same features as elliptic flow, i.e.\ the mass splitting and mass ordering
as expected from hydrodynamic models and a crossing point between pion and proton flow at
intermediate $p_t$ as expected from the quark coalescence picture.
Similarly to elliptic flow, triangular flow shows deviations from $KE_t$ scaling (see figure \ref{fig:figure5}).

\vspace{-0.3 cm}

\section{Summary}

We presented the $p_t$ differential elliptic flow of identified
particles for Pb-Pb collisions at $\sqrt{s_{NN}}=2.76\,\mathrm{TeV}$ measured with ALICE and
compared it to measurements at RHIC energies and a hydrodynamic model. 
The model correctly describes elliptic flow of pions and kaons,
but overpredicts the flow of protons for more central collisions. Compared to the RHIC data we
observed a larger mass splitting, mostly apparent in the proton flow.
We also showed deviations of elliptic flow from $KE_t$ scaling.
Additionally we presented the first measurement of p$_t$ differential triangular flow 
of identified particles at the LHC. We observed that $v_3$ has features similar to $v_2$, i.e.\ mass scaling 
and a crossing point for pions and protons at intermediate $p_t$.

\end{document}